**Evaluation of Dedicated Lanes for Automated vehicles at Roundabouts with Various Flow Patterns**

Amirreza Nickkar (Corresponding Author)
Ph.D. Student
Department of Transportation and Urban Infrastructure Studies
Morgan State University
1700 E Col Spring Lane
Baltimore, MD 21251
Tel: 443-885-3348; Fax:443-885-8275
Email: amnic1@morgan.edu;
ORCiD:0000-0002-1242-3778

Young-Jae Lee, Ph.D.
Associate Professor
Department of Transportation and Urban Infrastructure Studies
Morgan State University
1700 E. Cold Spring Lane, Baltimore, MD 21251
Tel: 443-885-1872; Fax: 443-885-8324
Email: YoungJae.Lee@morgan.edu
ORCiD: 0000-0002-1422-7965

Word Count: 3,149 words + 3 tables = 3,899 words

Nickkar and Lee

## ABSTRACT

Autonomous vehicles (AVs) are about to be used in transportation systems in the near future. To increase the level of safety and throughput of these vehicles, dedicated lanes for AVs have been suggested in past studies as exclusive mobility infrastructure for these types of vehicles. Although these lanes can bring obvious advantages in a transportation network, overall performance of these lanes, especially in urban areas and in micro level sight, is not clear. This study aims to examine the efficiency of dedicated lanes for AVs in a roundabout with an unbalanced traffic flow pattern. Four factors of travel time, delay time, speed of vehicles, and queue of vehicles have been selected as variables of traffic performance at the roundabout. Two microscopic traffic simulation software, AIMSUN and SIDRA Intersection, were used to examine the impact of the AV dedicated lanes at the roundabout. This study shows that the effects of an imbalanced traffic pattern in a roundabout are higher when the penetration rate of AVs is lower and also dedicated lanes in a roundabout's legs may improve traffic performance indicators when the penetration rate of AVs is higher, but this improvement is not significant.

# INTRODUCTION

The smart driving issue in transportation using automated facilities in urban roads, highways and freeways has been reviewed since the beginning of 1990s *(1)*. In recent years, a few studies have been conducted about dedicating special lanes to Automated Vehicles (AVs). AVs will have several advantages in future transportation network in terms of using minimum lane capacity, platooning of vehicles, safer lane changing behavior, less energy consumption, etc. *(2)*. The National Highway Traffic Safety Administration (NHTSA) defines vehicle automation as having levels 0, 1, 2, 3, and 4, which involve no automation, function-specific automation, combined-function automation, limited self-driving automation, and full self-driving automation, respectively *(3)*. AVs can run with minimum required spacing and headway, as opposed to no automation vehicles, and do so in mixed traffic conditions. Therefore, dedicating a number of lanes solely to AVs could enhance the performance of roads *(4)*.

Although dedicating lanes to AVs can bring some benefits, the effectiveness of assigning special lanes to AVs is still open to question because it depends on various factors such as penetration rate of AVs, automation level of AVs, traffic volume, type of the road, etc. Therefore, it's necessary to measure traffic indexes that show the performance of exclusive lanes for AVs in various traffic conditions. In recent years, many studies tried to understand the performance of AVs in the future transportation network by considering dedicated lanes and mixed traffic in simulation models. In future traffic networks, AVs would be allowed to drive in both mixed traffic and dedicated lanes. Taking a microscopic view of assessing the efficiency of dedicated lanes for AVs can give a better insight into the performance of these lanes in micro elements of a transportation network *(4)*.

In past decades, roundabouts benefitted the US as an alternative to signalized intersections because roundabouts have lower maintenance costs and less complexity than signalized intersections *(5)*. Although the performance of a roundabout in lowering traffic volumes and calming the traffic flow is remarkably better than an intersection, the efficiency of a roundabout is decreased by increasing traffic volume and especially discrepancy of traffic volume *(6)*. However, a study showed that utilizing ramp metering may improve performance of a roundabout under heavy traffic volume *(7)*, the performance of a roundabout after emerging AVs and especially under heavy traffic flow and unbalanced traffic patterns is still debatable. The current study aims to explore how dedicating a lane to automated vehicles when there is an unbalanced mix of AV and non-AV traffic patterns may affect traffic performance in the special case of a roundabout with an unbalanced traffic pattern. The results of this study can bring insights to predict the usefulness of implementing exclusive lanes for AVs in urban networks.

# LITERATURE REVIEW

Several studies have been conducted on how dedicated lanes for AVs impact the performance of highways and freeways in terms of some traffic features such as flow capacity, saturation flow, capacity drop and flow stability. Effects of Adaptive Cruise Control (ACC) and Cooperative Adaptive Cruise Control (CACC) have played a privileged role in describing exclusive lane efficiency *(8)*. The CACC uses a distance sensor to measure the distance and speed relative to vehicles driving ahead and is categorized as level 1 in the mentioned sorting. The issue of driving AVs in exclusive lanes is categorized as level 4 automation.

Some studies showed that AVs are able to triple roadway capacity. It has been proven that AVs, in terms of some special features such as platooning driving (vehicle groups traveling close together which reduces gap time), will positively impact the performance of the highway *(9)*. Some studies have demonstrated a positive influence of dedicated lanes for AVs. The dedicated lanes in the Automated Highway Systems (AHS) are able to increase safety and traffic flow *(10)*. Simulations have shown that fully automated vehicles in exclusive lanes could increase per-lane throughput by as much as three times, using inter-vehicle communications to exchange vehicle control parameters *(11)*. This finding shows similarities with the results of a study by Hall et al. *(12)* which demonstrated that increasing the number of AHS lanes beyond 2 or 3 only provides incremental capacity gains.

In contrast, some studies have challenged the positive effects of dedicated lanes for AVs. van der Werf et al. *(13)* assessed the effect of different penetration rates of CACC and ACC. They suggested that because spreading the ACC vehicles over different lanes leads to lessened shock waves, exclusive lanes are not profitable for ACC; conversely, dedicated lanes are able to increase the capacity when there is a high penetration rate of vehicles equipped with CACC. Similarly, van Arem et al *(14)* proposed a traffic flow simulation model MIXIC to assess the impact of CACC on the traffic flow. According to the results of this model, impacts of a dedicated lane for CACC-equipped vehicles have a direct relation with the CACC penetration rate. When there is a low level of CACC presence (<40%), a degradation of highway performance in terms of speed variation or creating shock wave is anticipated. On the other hand, when there is a high volume of traffic, the CACC lane performs better. Hussain et al. *(15)* concluded that as the CAV penetration rate increases, the freeway capacity increases as well and for the managed-lane problem, the decision on the lane allocation might take place only if the demand is greater than the freeway capacity. Also, Fakharian et al. *(16)* stated that that providing toll incentives for CACCs to use dedicated lanes is not beneficial at lower market penetration because of the small increase in capacity with these market penetrations. Such incentives are beneficial at higher market penetrations, particularly with higher demand levels.

The results of the studies outlined above suggest that dedicated lanes for passenger cars with ACC and CACC can have both a negative and a positive effect on the performance of a traffic network. Although some studies investigated performance of dedicated lanes for AVs in freeways and highways, there is a very little knowledge about the performance of dedicated lanes in roundabouts. To our knowledge, this is the first study to examine the efficiency of dedicated lanes for AVs in roundabouts with different traffic flow patterns.

## METHODOLOGY

The methodology of the current study is based on using microsimulation traffic software (AIMSUN 8.1) and micro-analytical traffic evaluation software (SIDRA Intersection 5.1) to investigate the performance of dedicated lanes for AVs on a roundabout with an unbalanced traffic flow pattern. Some studies examined the performance analysis of roundabouts under unbalanced traffic flow conditions. Akçelik *(17)* evaluated the performance of a roundabout with unbalanced traffic flow patterns. His research revealed that although an unbalanced traffic pattern in a roundabout when approaching traffic flows are low may not affect the performance of the roundabout, by increasing circulating traffic flow, the capacity of the roundabout would be

Nickkar and Lee

decreased due to a drop in acceptable gap rates in approaching flow. Krogscheepers and Roebuck (18) showed that by increasing imbalance ratio of circulating traffic flow of a roundabout the delay time will be increased and the delay time is highly dependent on the origin of conflicting traffic. This study used the roundabout case study of Krogscheepers and Roebuck *(18)* data set and roundabout attributes. The test roundabout is four-legged with two driving lane in each approach (Figure 1). The roundabout has an island that is 30 meters in diameter and 10 meters circulating width. Each leg of the roundabout is 500 meters long with two lanes. The data set has two main features: variation of the conflicting circulating traffic stream and avoiding over-saturation traffic flow condition in the roundabout (Table 1). The Ratio of Imbalance (Rho) has been determined to adjust the effect of imbalanced traffic patterns in the roundabout. The definition of this ratio is according to Krogscheepers and Roebuck *(18)* study. They defined this ratio as "*the proportion of circulating traffic originating from the first upstream approach of the subject approach*" which means for example for the northern approach the Rho will be as equition (1):

$$\text{Rho}_N = \frac{R_E}{T_{WN} + T_{SN} + R_E} \tag{1}$$

Where, $T_{WN}$ is traffic flow from west passing to north, $T_{SN}$ is traffic flow from south to north, and $R_E$ is traffic flow from east to north. The Rho ranges from 0 to 1 in intervals of 0.1 and in 11 various scenarios. The results of the micro-analysis model show that unbalanced traffic flows for conventional vehicles in a roundabout affect all traffic performance indexes of delay time (sec) and speed (km/hr) (Table 2 and Figure 2). In this study two differences cases for the roundabout has been introduced; without a dedicated lane in approaching roads, and with a dedicated lane in all approaching roads. Also, three different cases as 25%, 50%, and 75% have been considered as the penetration rates of AVs and finally 11 different scenarios for evaluating the effect of imbalance traffic flow patterns therefore, totally 66 different scenarios have been considered in the current study.

**Microsimulation analysis**

In this section, two microsimulation traffic models of the roundabout have been conducted using AIMSU. The first model details the results of the impact of the presence of AVs as a part of unbalanced traffic flow pattern on the performance of the roundabout. In the second model, it is assumed that one lane in each leg has been dedicated to the AVs to evaluate performance of these lanes in mixed traffic conditions in a roundabout with an unbalanced traffic flow pattern. Two types of human-driven vehicles, (HVs) and AVs, have been introduced in the model, and the AV penetration rate varied between 25% and 75% in intervals of 25% to examine the effects that increased AV penetration rate has on the performance of a roundabout with an unbalanced traffic flow pattern.

Nickkar and Lee

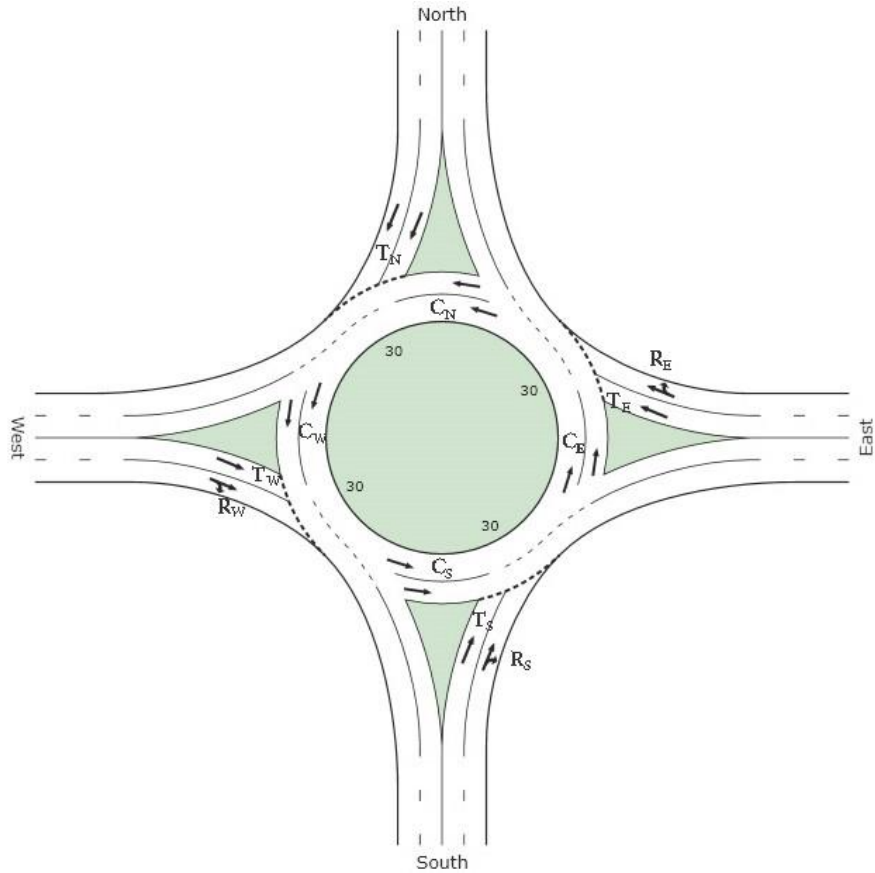

**Figure 1. The schematic of a roundabout**

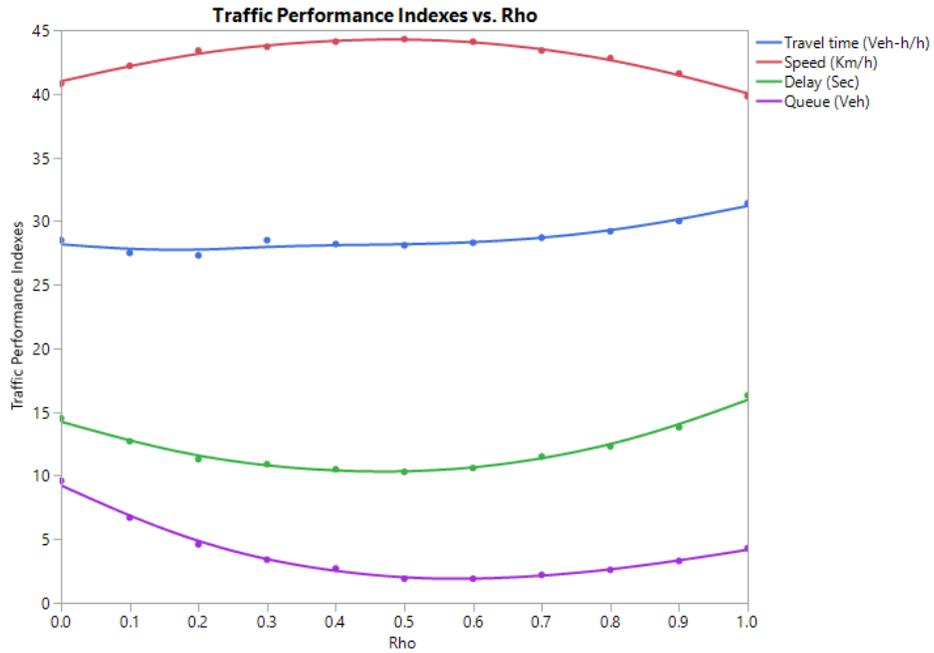

**Figure 2. Traffic performance indexes by variation of Rho**

**Table 1. Data set in micro-analytical and microsimulation analysis (18)**

| Scenario | North | | | East | | | South | | | West | | | Total | |
|---|---|---|---|---|---|---|---|---|---|---|---|---|---|---|
| | $T_N$ | Rho | $C_N$ | $T_E^1$ | $R_E^2$ | $C_E^3$ | $T_S$ | $R_S$ | $C_S$ | $T_W$ | $R_W$ | $C_W$ | $TF^4$ | $TCF^5$ |
| 1 | 600 | 1 | 11 | 10 | 300 | 239 | 220 | 10 | 380 | 350 | 340 | 652 | 2449 | 1282 |
| 2 | 600 | 0.9 | 11 | 10 | 300 | 239 | 220 | 80 | 337 | 310 | 310 | 652 | 2406 | 1239 |
| 3 | 600 | 0.8 | 11 | 10 | 320 | 217 | 200 | 150 | 326 | 300 | 250 | 652 | 2373 | 1206 |
| 4 | 600 | 0.7 | 11 | 10 | 340 | 196 | 180 | 200 | 304 | 280 | 220 | 652 | 2330 | 1163 |
| 5 | 600 | 0.6 | 11 | 10 | 340 | 196 | 180 | 280 | 239 | 220 | 200 | 652 | 2265 | 1098 |
| 6 | 600 | 0.5 | 11 | 10 | 340 | 196 | 180 | 340 | 196 | 180 | 180 | 652 | 2222 | 1055 |
| 7 | 600 | 0.4 | 11 | 10 | 390 | 141 | 130 | 420 | 152 | 140 | 140 | 652 | 2123 | 956 |
| 8 | 600 | 0.3 | 11 | 10 | 420 | 109 | 100 | 480 | 120 | 110 | 110 | 652 | 2059 | 892 |
| 9 | 600 | 0.2 | 11 | 10 | 350 | 87 | 80 | 550 | 87 | 80 | 70 | 652 | 1914 | 837 |
| 10 | 600 | 0.1 | 11 | 10 | 350 | 54 | 50 | 630 | 33 | 30 | 40 | 652 | 1797 | 750 |
| 11 | 600 | 0 | 11 | 10 | 350 | 54 | 50 | 690 | 5 | 5 | 5 | 652 | 1769 | 722 |

[1]Through flow from East approach; [2]Right flow from East approach; [3]Circulating flow in the East side of the roundabout; [4]Total flow; [5]Total circulating flow

**Table 2. Results of micro-analysis model of the roundabout**

| Scenario | North | | | | East | | | | South | | | | West | | | | Total | | | |
|---|---|---|---|---|---|---|---|---|---|---|---|---|---|---|---|---|---|---|---|---|
| | $TT_N$ | $S_N$ | $D_N$ | $Q_N$ | $TT_E$ | $S_E$ | $D_E$ | $Q_E$ | $TT_S$ | $S_S$ | $D_S$ | $Q_S$ | $TT_W$ | $S_W$ | $D_W$ | $Q_W$ | $TT_T^1$ | $S_T^2$ | $D_T^3$ | $Q_T^4$ |
| 1 | 8.7 | 47 | 8 | 1.4 | 4.8 | 43.4 | 10.7 | 1.6 | 3.3 | 47.1 | 7.9 | 0.6 | 14.4 | 32.5 | 28.8 | 4.3 | 31.4 | 39.8 | 16.3 | 4.3 |
| 2 | 8.7 | 47 | 8 | 1.4 | 4.8 | 43.4 | 10.7 | 1.6 | 4.4 | 46.4 | 8.3 | 0.7 | 12 | 35.1 | 23.7 | 3.3 | 30 | 41.6 | 13.8 | 3.3 |
| 3 | 8.7 | 47 | 8 | 1.4 | 5.2 | 43.4 | 11 | 1.8 | 5.2 | 45.8 | 8.8 | 0.9 | 10 | 37.3 | 20 | 2.6 | 29.2 | 42.8 | 12.3 | 2.6 |
| 4 | 8.7 | 47 | 8 | 1.4 | 5.5 | 43 | 11.2 | 1.9 | 5.7 | 45.5 | 8.9 | 1 | 8.8 | 38.6 | 17.9 | 2.2 | 28.7 | 43.4 | 11.5 | 2.2 |
| 5 | 8.7 | 47 | 8 | 1.4 | 5.5 | 43 | 11.2 | 1.9 | 7 | 44.9 | 9.4 | 1.5 | 7.1 | 40.5 | 15.2 | 1.7 | 28.3 | 44.1 | 10.6 | 1.9 |
| 6 | 8.7 | 47 | 8 | 1.4 | 5.5 | 43 | 11.2 | 1.9 | 8 | 44.3 | 10.1 | 1.9 | 5.9 | 41.7 | 13.5 | 1.3 | 28.1 | 44.3 | 10.3 | 1.9 |
| 7 | 8.7 | 47 | 8 | 1.4 | 6.4 | 42.5 | 11.8 | 2.3 | 8.7 | 42.9 | 11.6 | 2.7 | 4.4 | 43.1 | 11.7 | 0.9 | 28.2 | 44.1 | 10.5 | 2.7 |
| 8 | 8.7 | 47 | 8 | 1.4 | 6.9 | 42.3 | 12.1 | 2.6 | 9.5 | 41.5 | 13.3 | 3.4 | 3.4 | 44.1 | 10.5 | 0.7 | 28.5 | 43.7 | 10.9 | 3.4 |
| 9 | 8.7 | 47 | 8 | 1.4 | 5.5 | 44.2 | 9.8 | 1.8 | 10.8 | 39.6 | 15.8 | 4.6 | 2.3 | 45.2 | 9.3 | 0.4 | 27.3 | 43.4 | 11.3 | 4.6 |
| 10 | 8.7 | 47 | 8 | 1.4 | 5.5 | 44.6 | 9.3 | 1.7 | 12.3 | 37.5 | 19 | 6.7 | 1 | 46.2 | 8.1 | 0.2 | 27.5 | 42.2 | 12.7 | 6.7 |
| 11 | 8.7 | 47 | 8 | 1.4 | 5.5 | 44.6 | 9.3 | 1.7 | 14.1 | 35.5 | 22.3 | 9.6 | 0.1 | 47.1 | 7.3 | 0 | 28.5 | 40.8 | 14.5 | 9.6 |

[1] Total travel time (veh-h/h); [2]Speed (km/h); [3]Delay (sec); [4]Queue (veh)

The class of AVs should be designed and calibrated in the simulation software because behavioral characteristics of AVs depend on the automation level of these vehicles, which are different from HVs. The class of AVs in this study is coordinated to level 3 of automation according to NHTSA's vehicle automation classification. In level 3, an Automated Driving System (ADS) controls steering and braking/accelerating and all other related driving tasks in a certain driving circumstance. But the ADS is not able cover all tasks of driving and the driver must be ready to control the vehicle at any time and pay attention when circumstances change, and vehicle drivers are in an environment with no connectivity with other vehicles. Bailey *(19)* calibrated this level of automation in AIMSUN by calibrating acceleration/breaking, car following and gap acceptance behaviors. In this study, some new parameters for the specific case of a roundabout have been calibrated to improve the accuracy of the simulation models. In this study, it is assumed that the left lane of each leg of the roundabout has been dedicated for AVs.

**Table 3. Recommended calibrated parameters for AVs in AIMSUN** *(19, 20)*

| Parameter | HV (Mean) | AV |
|---|---|---|
| Speed Acceptance | 1.1 | 1.1 |
| Minimum Gap (m) | 1 | 1 |
| Lane-Changing Cooperation | Not Checked | Checked |
| Imprudent Lane-Changing | Yes | Not |
| Initial Safety Margin (s) | 3 | 1 |
| Final Safety Margin (s) | 1 | 1 |
| Initial Give-Way Time Factor (s) | 1 | 1 |
| Reaction Time (s) | 0.8 | 0.1 |
| Reaction Time at Stop (s) | 1.2 | 0.1 |
| Max Acceleration (m/s$^2$) | 3 | 3 |
| Max Deceleration (m/s$^2$) | 6 | 6 |

To investigate the efficiency of dedicated lanes in a roundabout with an unbalanced traffic pattern, two modeling approaches – with and without dedicated AV lanes – have been used. In each approach, 11 Rhos scenarios (presented in Table 1) with three levels of penetration rate of AVs (AVPR) were combined, and a total of 66 scenarios have been conducted. In the first approach, it was assumed that there is no dedicated lane for AVs in a roundabout's legs and the traffic flow of AVs and HVs was mixed. The results of the simulation models in this approach show that traffic performance indexes of the roundabout when the penetration rate of AVs is higher indicate better results. Also, by increasing unbalanced traffic flow in the roundabout, the scenario with 25% AVPR is impacted more than the scenario with 75% AVPR (Figure 3).

In the second approach, it is assumed that the left lane in each leg of the roundabout is dedicated for AVs. The results of simulation models in this approach show that the performance of the roundabout in lower AVPR is noticeably reduced. However, the performance of the roundabout in the higher AVPR is slightly better than the case of the first approach.

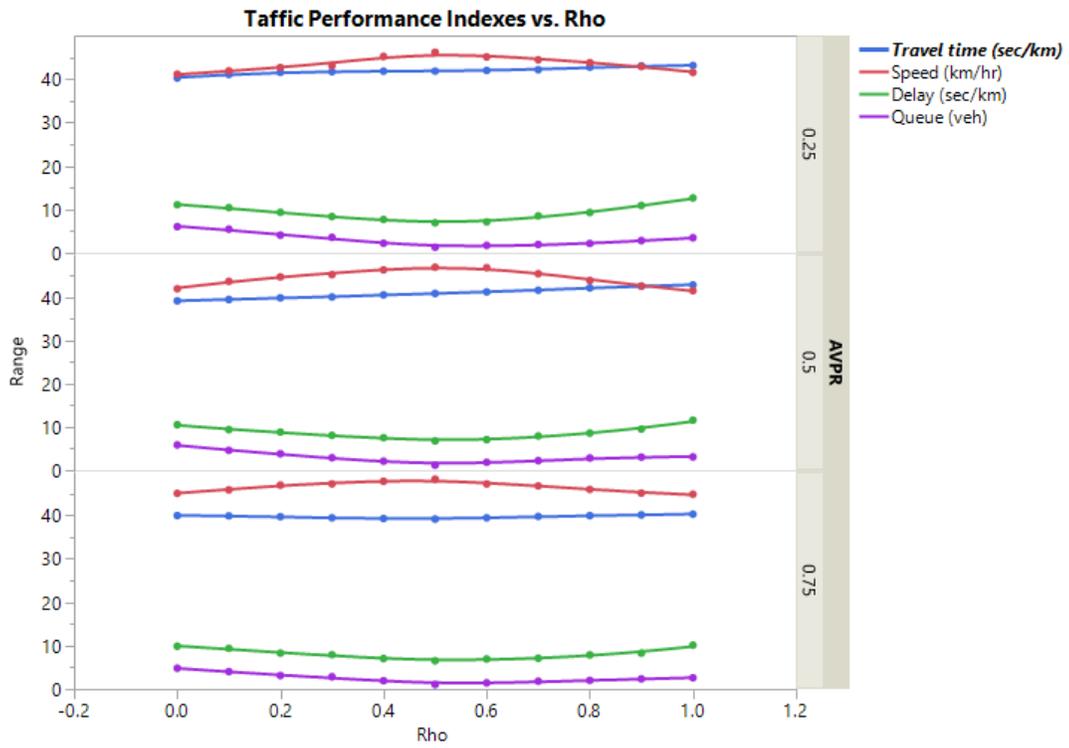

**Figure 3. Effect of unbalanced traffic flow in a roundabout without dedicated lanes for AVs**

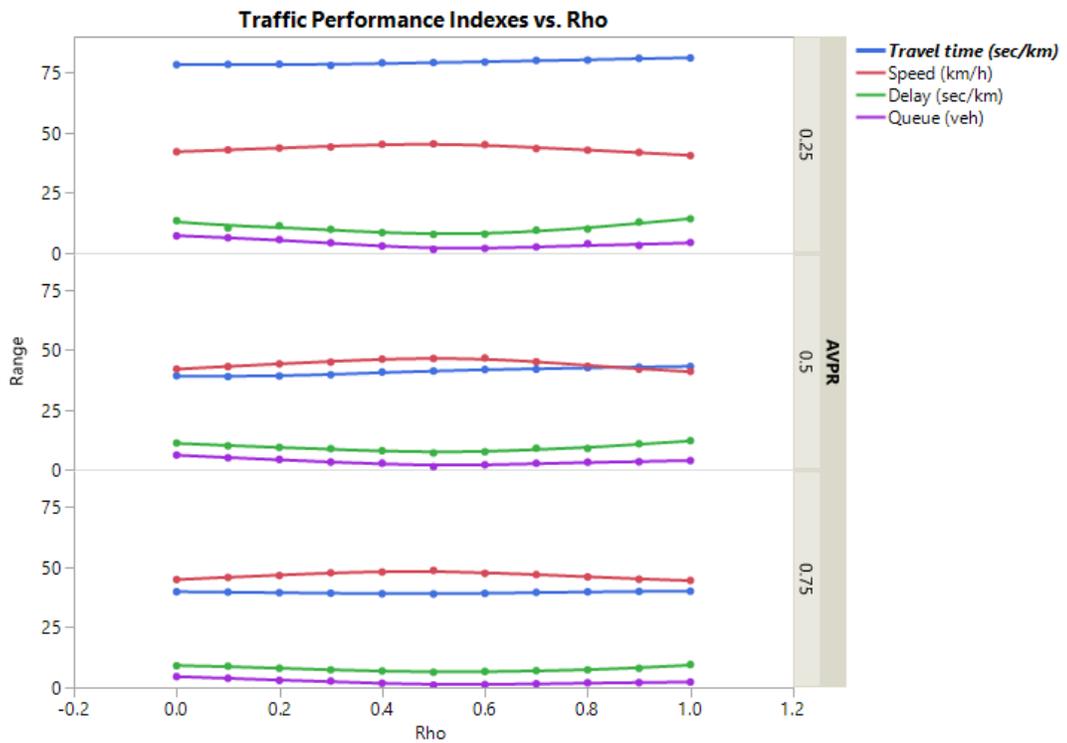

**Figure 4. Effect of an unbalanced traffic flow in a roundabout with dedicated lanes for AVs**

**CONCLUSION**

This paper examined the efficiency of dedicated lanes for AVs in a roundabout with various traffic flow patterns. The results of this study provide two main findings: first, when the penetration rate of AVs is low, the effect of imbalanced traffic in the roundabout with a dedicated lane for AVs would be higher than that without a dedicated lane for AVs; second, although dedicating a lane for AVs improves traffic performance indexes in scenario with 75% AVPR, this improvement is not significant compared to the case without a dedicated lane for AVs. Also, dedicating lanes for AVs when the penetration rate of AVs is low in a roundabout would decrease the efficiency of the roundabout. The results show that as the AV penetration rate increases in the roundabout without a dedicated lane for AVs, the travel time of vehicles is reduced because of better driving performance parameters of AVs than those of conventional vehicles, therefore, the effect of an imbalanced traffic pattern with a higher AV penetration rate is less than that with a lower AV penetration rate,

Nevertheless, implementing a dedicated lane directly reduced the road capacity when the AV penetration rate is low. In condition of lower penetration rate of AVs, the effect of an imbalanced traffic pattern is remarkably higher even compared with those scenarios with all conventional vehicles. Moreover, in condition of higher penetration rate of AVs when there is a dedicated lane for AVs, reduction of travel time and delay was not significant especially in scenarios with higher imbalanced traffic patterns.

The future research can use fully automated vehicles and connected automated vehicles in simulation modeling. Moreover, considering three-lane roundabouts in simulation modeling could be another extension of this study.

**AUTHOR CONTRIBUTION STATEMENT**

The authors confirm contribution to the paper as follows: study conception and design; data collection; analysis and interpretation of results; draft manuscript preparation: Amirreza Nickkar, Young-Jae Lee. All authors reviewed the results and approved the final version of the manuscript.